\documentclass[10pt,pra,aps,amsmath,amssymb,amsfonts,twocolumn,nofootinbib,longbibliography,superscriptaddress]{revtex4-2}
\usepackage{amssymb}
\usepackage{bm,mathrsfs}
\usepackage{graphicx}
\usepackage{subfigure}
\usepackage{epsfig}
\usepackage{amsmath,bbm}
\usepackage{amsfonts,amssymb}
\usepackage{times}
\usepackage{verbatim}
\usepackage[sort&compress]{natbib}
\usepackage{amsmath}
\usepackage{microtype}  
\usepackage[colorlinks=true,citecolor=blue,linkcolor=blue,urlcolor=blue]{hyperref}
\usepackage[usenames]{color}

\newcommand{\bra}[1]{\langle #1|}
\newcommand{\ket}[1]{|#1\rangle}

\newcommand{\figpanel}[2]{\hyperref[#1]{\ref*{#1}(#2)}}

\begin{document}

\title{Non-Markovian giant-atom dynamics in a disordered lattice}

\author{Maohua Wang}
\affiliation{Center for Quantum Sciences and School of Physics, Northeast Normal University, Changchun 130024, China}

\author{Yan Zhang}
\email{zhangy345@nenu.edu.cn}
\affiliation{Center for Quantum Sciences and School of Physics, Northeast Normal University, Changchun 130024, China}

\date{\today}
\begin{abstract}
While ideal lattice models have been widely used to study giant-atom systems, fabrication-induced defects inevitably introduce disorder in realistic platforms. 
Here, we study non-Markovian dynamics of a giant atom coupled to a discrete photonic lattice with on-site frequency disorder. 
Using time-domain and spectral analyses, we show that the overall population-decay envelope and global photon-transport patterns remain robust against moderate lattice disorder, while the quantified non-Markovian memory can be significantly enhanced within the explored disorder range.
We characterize the memory using a normalized geometrical non-Markovianity measure tailored to delayed giant-atom feedback and demonstrate how the coupling-point separation and the disorder strength serve as complementary parameters that shape the delay timescale and the complexity of coherent-feedback interference.
Spectral analysis reveals that scattering-band transport is relatively insensitive to disorder, whereas disorder-sensitive bound-state branches and localization features reshape revival windows and promote information backflow. Our results establish a disorder-aware framework for understanding and engineering non-Markovian feedback effects of giant atoms in structured reservoirs.

\end{abstract}

\maketitle

\section{Introduction}
\label{SecI}

Understanding light-matter interactions has long been central to quantum optics and has stimulated extensive research on both natural and artificial atomic systems. 
Artificial atoms, such as quantum dots~\cite{qd1} and superconducting qubits~\cite{sq1,sq2,sq3,sq4,sq5}, are engineered to exhibit discrete energy-level structures analogous to those of natural atoms. 
Within the electric-dipole approximation, these systems are typically treated as point-like ``small atoms", since their physical dimensions are much smaller than the wavelength of the relevant electromagnetic fields.
In recent years, giant atoms, quantum emitters that couple nonlocally to their environment through multiple spatially separated points, have emerged as a novel framework in quantum optics that goes beyond the conventional dipole approximation~\cite{ls1, PhysRevA2017.053821, PRA2020.101.053855, kockum2021fiveyear, PRL2022.128.223602, du2023giant}. 
The nonlocal coupling renders the dynamics sensitive to the phase accumulated by the mediating field between different coupling points, giving rise to the hallmark feature of giant atoms, the self-interference effect. 
Giant-atom physics has been explored in a variety of platforms, including superconducting quantum circuits~\cite{sqc1,sqc2}, coupled-waveguide arrays~\cite{wa1,wa2}, natural-atom implementations~\cite{ra1,ra2,ra3,ra4}, and spin ensembles~\cite{se1}. 
Owing to their extended geometry, the resulting self-interference effect leads to a range of distinctive phenomena, such as frequency-dependent relaxation and Lamb shifts~\cite{ls1,ls2}, non-exponential decay~\cite{ned1,ned2,ned3,ned4,ned5}, decoherence-free interactions~\cite{dfi1,dfi2,dfi3,dfi4}, novel bound states~\cite{obs1,obs2,obs3,obs4,obs5,obs6,obs7}, controllable frequency conversion~\cite{fc2,fc3,fc4}, and long-lived entanglement~\cite{qe1,qe2,qe3,qe4,qe5,du2025dressed}.


A key consequence of the spatial extent of giant atoms is the emergence of non-Markovian retardation effects~\cite{nmm1, LeiDu12301, PRA2023.107.023716, PRL2024.133.063603}. 
In contrast to small atoms, where propagation times are typically negligible, giant atoms experience finite propagation delays even in the weak-coupling regime~\cite{nm2,nm3,nm4,nm5,nm6,nm7,nm8,nm9,nm10}. 
When the propagation time $\tau$ between coupling points is much shorter than the atomic relaxation time $\Gamma^{-1}$, the dynamics remain effectively Markovian. 
As $\tau$ becomes comparable to $\Gamma^{-1}$, however, the system enters a non-Markovian regime in which photons emitted from one coupling point can be reabsorbed at another, leading to information backflow---a hallmark of non-Markovian dynamics~\cite{PRR2022.4.023198, Seyed24201}. 
Such delay-induced phenomena have been demonstrated in a variety of quantum platforms, including large geometric atoms~\cite{ned1}, emitters near mirrors~\cite{nmde2}, cavity–giant-atom configurations~\cite{nmde3}, macroscopically separated emitters~\cite{nmde4}, and chains of distant qubits~\cite{nmde5}. 
Numerical and analytical approaches to these systems include matrix-product-state simulations~\cite{mps}, discretized waveguide models~\cite{dwm}, and diagrammatic techniques~\cite{da}. 
Prominent non-Markovian effects include unconventional oscillatory bound states supported by multiple coupling points~\cite{osbs1} and a variety of interference-enabled chiral behaviors, such as nonreciprocal frequency conversion~\cite{frc}, dark states without external driving~\cite{ds}, spontaneous chiral emission~\cite{sqc2}, and port-selective routers and circulators~\cite{cir}. 

To gain deeper insight into the physical phenomena arising from the non-Markovian properties of a giant atom coupled to a lattice, it is essential to account for structural defects that are inevitably introduced during lattice fabrication, resulting in lattice disorder. 
Disorder in giant-atom platforms can also arise from imperfect couplings themselves, such as fluctuations in coupling strengths or uncertainties in coupling positions, and these imperfections have been shown to modify robustness and non-Markovian signatures in a disorder-dependent manner~\cite{Guo2026Disorder}. 
Here, we focus on disorder in the structured reservoir, namely on-site frequency fluctuations in a discrete lattice, and analyze how such lattice disorder reshapes delayed feedback, spectral features, and the quantified non-Markovian memory. 
In experiments, on-site disorder may arise unintentionally from fabrication imperfections, but it can also be engineered in a controlled statistical manner, for example, by calibrated femtosecond-laser writing variations that set propagation-constant detunings or by introducing controlled disordered potentials in ultracold-atom platforms~\cite{Billy2008Nature, Roati2008Nature, Martin2011OE, Biggerstaff2016NatCommun}. 
We therefore treat the disorder strength as an effective statistical parameter and scan it to quantify which observables are disorder-insensitive and how non-Markovian memory responds.

In this work, we present a disorder-aware study of non-Markovian giant-atom lattice dynamics. 
First, we provide a systematic time-domain analysis of atomic population and photon transport under on-site frequency disorder, identifying a robustness window in which global dynamics remain stable. 
Then, we use a normalized non-Markovianity metric tailored to delayed giant-atom feedback and show that disorder can serve as an active resource to enhance information backflow. 
Third, we connect time-domain revivals to spectral features and establish a unified physical picture. 
Scattering-band transport remains relatively robust, whereas bound-state branches are fluctuation-sensitive in the disordered lattice.
These findings bridge realistic fabrication imperfections and controllable non-Markovian engineering in giant-atom platforms.


\section{Model and Equations}

\label{SecII}
\begin{figure}[tpb]
\includegraphics[width=8.4 cm]{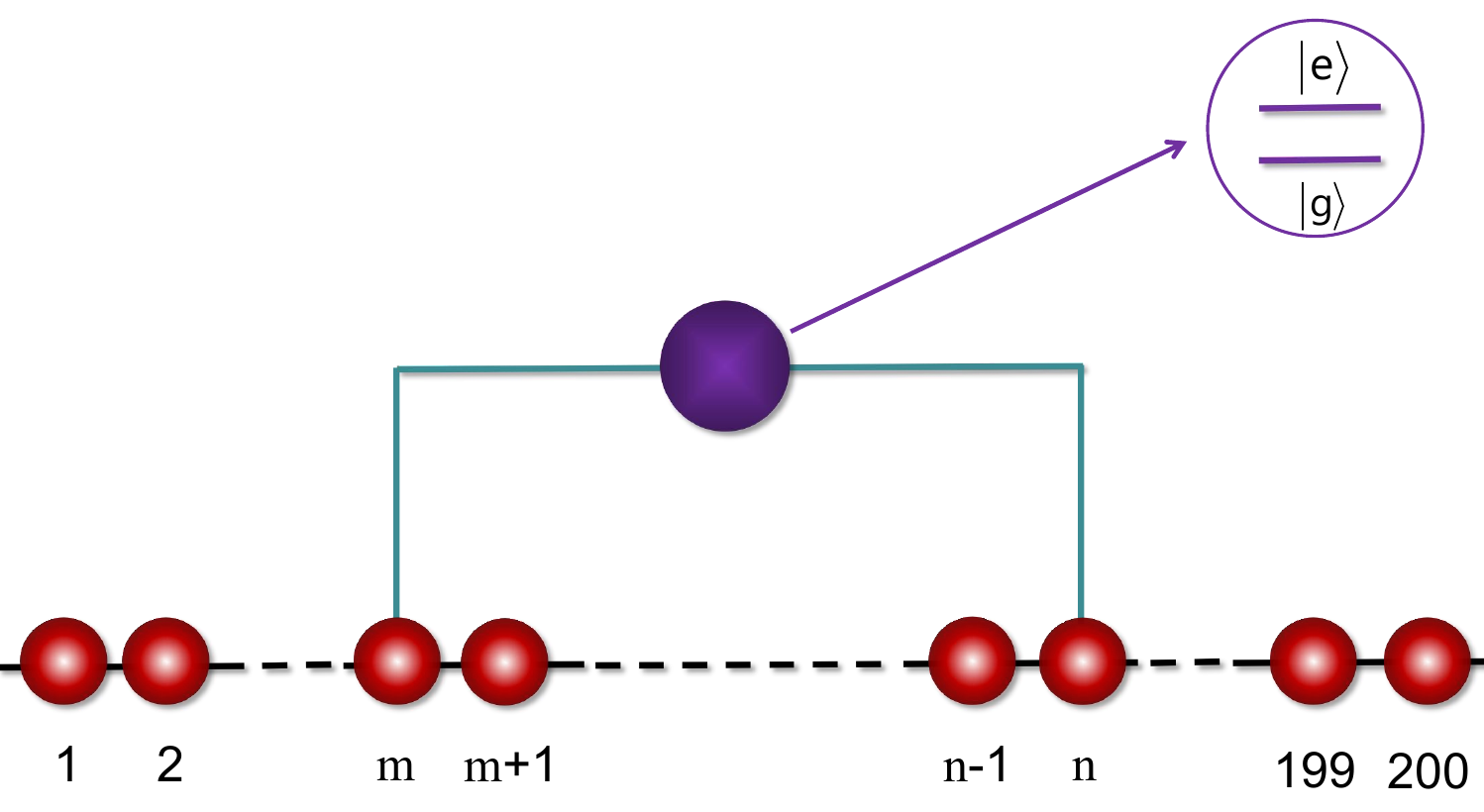}
  \caption{Schematic configuration for a discrete lattice coupled to a giant atom at the $m$ site and $n$ site.}
\label{F1}
\end{figure}

As shown in Fig.~\ref{F1}, the giant atom system with two coupling points is composed of a one-dimensional discrete-waveguide lattice with $200$ sites and an emitter with a ground state $\ket{g}$ and excited state $\ket{e}$. 
The system Hamiltonian $H$ is divided into three parts, i.e., $H=H_{0}+H_{j}+H_{I}$. 
$H_{0}$ is the free Hamiltonian of the giant atom as ($\hbar$ = 1)
\begin{equation}\label{eq_01}
H_{0}= \omega_{e}\ket{e} \bra{e},
\end{equation}
where $w_{e}$ is the transition frequency between states $\ket{e}$ and $\ket{g}$, and we set the energy of the ground state $\ket{g}$ to $\omega_{g} = 0$.
$H_{j}$, representing the free Hamiltonian of the lattice, is expressed as
\begin{equation}\label{eq_02}
{H}_{j}=\sum_{j}(\omega_{0}+\delta_{j}){a}_{j}^\dag{a}_{j}-J({a}_{j}^\dag{a}_{j+1}+{a}_{j+1}^\dag{a}_{j}),
\end{equation}
where $\omega_0$ represents the frequency of the waveguide mode, $J$ is the hopping strength between the nearest lattice sites, $\delta_{j}$ denotes the frequency fluctuation and the degree of disorder at the $j$ site, and $a_{j}\ (a_{j}^\dag)$ is the bosonic annihilation (creation) operator at the site $j$.
$H_{I}$ describes the interaction between the lattices and the giant atom. 
Under the rotating wave approximation, $H_{I}$ is expressed as
\begin{equation}\label{eq_03}
H_{I}=(g_{m}a_{m}^\dag+g_{n}a_{n}^\dag)\sigma^{-}+(g_{m}a_{m}+g_{n}a_{n})\sigma^{+},
\end{equation}
where $\sigma^{+}=(\sigma^{-})^{\dag}=\ket{e} \bra{g}$ are Pauli operators of the giant atom.
Here, we have considered the giant atom coupled to the lattice at the $m$ site and $n$ site with coupling strengths $g_{m}$ and $g_{n}$, respectively. 

Restricting to the single-excitation subspace, we express the system wave function as
\begin{equation}\label{eq_04}
\ket{\psi(t)}=[\sum_{j}{C_{j}(t)a_{j}^{\dag}}+C_{e}(t)\sigma^{+}]\ket{0,g},
\end{equation}
where \(C_j(t)\) denotes the probability amplitude for a photon at site \(j\), \(C_e(t)\) represents the excitation amplitude of the giant atom, and \(\ket{0,g}\) describes the lattice in the vacuum state and the giant atom in its ground state \(\ket{g}\).

The system dynamics is governed by the Schr$\ddot{o}$dinger equation $i\ket{\dot{\psi}(t)}=H\ket{\psi(t)}$. Then, we obtain the equations of motion of a giant atom and the time evolution of the lattices, described as
\begin{equation}\label{eq_05}
\begin{split}
i\dot{C}_{j}(t)=&(\omega_{0}+\delta_{j})C_{j}(t)-JC_{j+1}(t)-JC_{j-1}(t)\\
&+g_{m}C_{e}(t)\delta_{j,m}+g_{n}C_{e}(t)\delta_{j,n},\\
i\dot{C}_{e}(t)=&\omega_{e}C_{e}(t)+g_{m}C_{m}(t)+g_{n}C_{n}(t).
\end{split}
\end{equation}

\section{Results and discussion}
\label{SecIII}

Conventional studies of atom-lattice coupling typically assume an ideal lattice, even though experiments reveal that structural defects and disorder are unavoidable in realistic optical lattice platforms. 
Building on this motivation, we present a disorder-aware study of giant-atom lattice dynamics, systematically quantifying how on-site disorder modifies population decay, photon transport, and non-Markovian information backflow.

\subsection{Effects of disorder on giant-atom evolution and non-Markovian dynamics}
\label{SecIIIA}

To quantify quantum non-Markovianity in open systems, several theoretical measures have been proposed. 
In this work, we use the measure $N$ that is defined in terms of the volume of accessible states~\cite{nmm1,nmm2}. 
The degree of quantum non-Markovianity is quantified by 
\begin{equation}\label{eq_12}
N_{V}= \int_{ \partial{t\left| C_{e}(t) \right|>0}}dt\frac{d\left| C_{e}(t) \right|^{4}}{dt}.
\end{equation}
Non-Markovian behavior occurs whenever $\partial_{t}\left| C_{e}(t) \right|>0$, that is, when the excitation probability of the giant atom increases over time. 
These revival windows provide a direct operational signature of backflow in the single-excitation subspace.
Consequently, $N_{V}>0$ if and only if $\left| C_{e}(t) \right|$ exhibits increases at certain times, whereas a purely monotonic decay of \(\left| C_{e}(t) \right|\) implies \(N_{V}=0\), corresponding to Markovian dynamics.

To account for regimes in which stronger atom-environment coupling gives rise to vacuum Rabi oscillations, we introduce a normalized (rescaled) non-Markovianity measure.
Assuming $\left| C_{e}(0) \right| = 1$ and $\left| C_{e}(\infty) \right| = 0$, we define
\begin{equation}\label{eq_13}
\begin{split}
N &= \frac{N_{V}}{\left|\int_{\partial_{t}\left| C_{e}(t) \right|<0} dt\,\frac{d\left| C_{e}(t) \right|^{4}}{dt}\right|}
= \frac{N_{V}}{N_{V}-\int_{0}^{\infty} dt\,\partial_{t}\left| C_{e}(t) \right|^{4}} \\
&= \frac{N_{V}}{N_{V}+1}.
\end{split}
\end{equation}
By construction, for purely Markovian dynamics $\left| C_{e}(t) \right|$ decreases monotonically in time, implying $N_{V}=0$ and hence $N=0$.

\begin{figure}[t]
\includegraphics[width=8.4 cm]{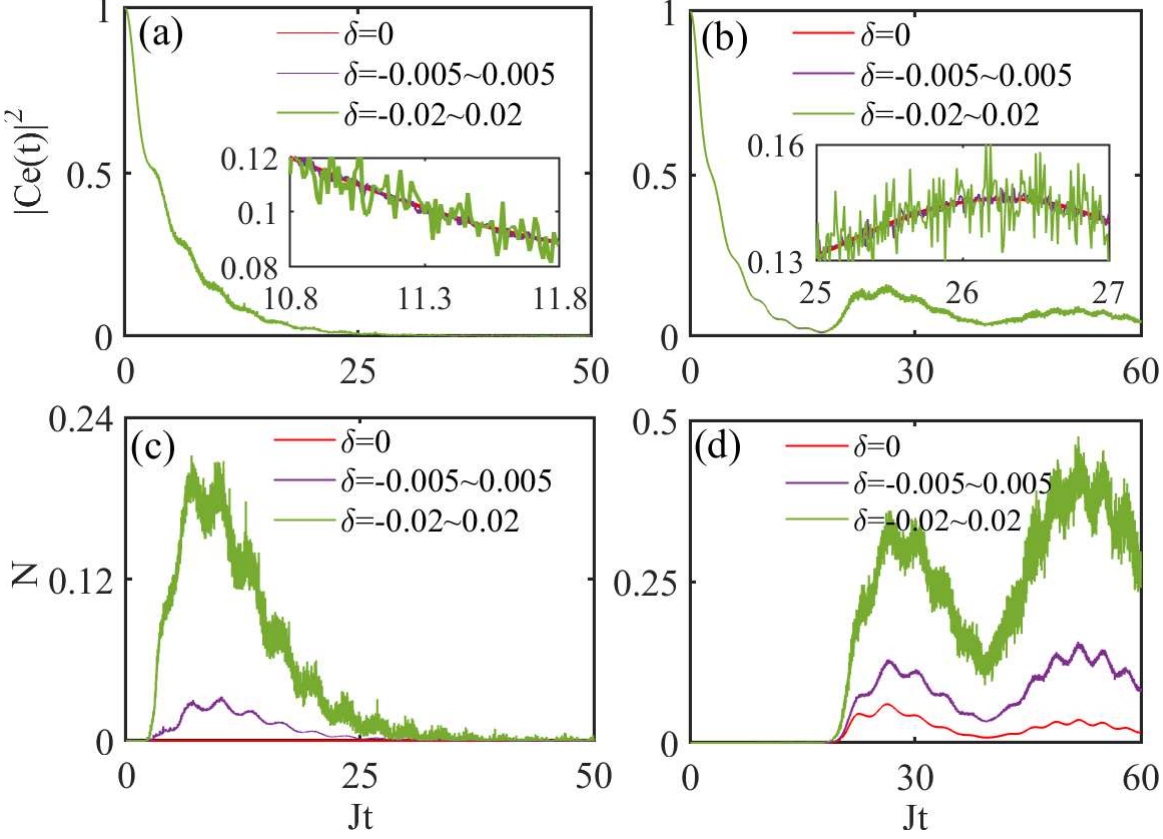}
\caption{(a) Evolution of the excited-state population $|C_e(t)|^2$ and (c) non-Markovianity measure $N$ for a giant atom with coupling points at $m=99$ site and $n=102$ site.
(b) Evolution of the excited-state population $|C_e(t)|^2$ and (d) non-Markovianity measure $N$ for a giant atom with coupling points at $m=83$ site and $n=118$ site.
The red, purple, and green solid curves correspond to disorder parameters $\delta=0$, $\delta \in [-0.005,0.005]$, and $\delta \in [-0.02,0.02]$, respectively.
The insets in (a) and (b) show enlarged views that allow the three curves to be distinguished more clearly.
Other parameters are $\omega_{0}=2J$, $\omega_{e}=2J$, and $g_{m}=g_{n}=0.35J$.
}
\label{F2}
\end{figure}

Figures~\figpanel{F2}{a} and \figpanel{F2}{b} illustrate the dynamical evolution of the excited-state population of the giant atom under different disorder strengths. 
For the configuration with coupling points at $m=99$ and $n=102$ [see Fig.~\figpanel{F2}{a}], the decay of the atomic excitation remains relatively stable and follows a similar trend as the disorder strength $\delta$ is increased from $0$ to $0.02$. 
In the absence of disorder, the excitation decays smoothly and approximately exponentially, with small oscillations originating from giant-atom self-interference. 
Owing to the short coupling-point separation $|m-n|$ (the separation between the two coupling points), information backflow sets in before the population has completely relaxed to the ground state. 
As the disorder strength increases, the decay acquires noticeable fluctuations.

When the coupling-point separation $|m-n|$ is increased, as shown in Fig.~\figpanel{F2}{b}, the system exhibits qualitatively different behavior. 
For sufficiently large $|m-n|$, a reflux of information appears only after the excited-state population has essentially decayed to zero. 
In the early-time interval $Jt\in[0,20]$, the excited-state population decays almost exponentially, and no re-excitation events occur, corresponding to conventional spontaneous-emission dynamics. 
In the later-time interval $Jt\in[20,40]$, the atom becomes re-excited by absorbing delayed photons emitted from the other coupling site, leading to periodic revivals of its quantum state. 
Even in the disordered case, the decay profile remains structurally robust; disorder primarily broadens revival features rather than changing the dominant relaxation channel.

This behavior can be understood from the interplay between spectral structure and delayed coherent feedback in a disordered lattice. 
In the large-$|m-n|$ regime, the non-Markovianity mainly originates from delayed backflow. 
Photons emitted into the waveguide can return to the atom after a finite flight time, generating revival windows in $|C_e(t)|$. 
Spectrally, the scattering band provides the dominant extended transport channel that governs the overall relaxation envelope and remains relatively robust against moderate on-site disorder. 
By contrast, disorder strongly perturbs the bound-state branch and induces localization features, which enrich the coherent-feedback pathways and introduce site-dependent phase shifts and scattering-induced interference. As a result, the returning field becomes temporally more spread and irregular, so the revival windows are broadened rather than replaced by a new dominant decay channel. Since the normalized non-Markovianity $N$ integrates the growth segments where $\partial_{t}\left|C_{e}(t) \right|>0$, these broadened and irregular revival windows accumulate more backflow, leading to an enhanced $N$ within the explored disorder window.

Figures~\figpanel{F2}{c} and~\figpanel{F2}{d} present the dynamics of the non-Markovianity measure $N$. 
The results show a monotonic enhancement trend of $N$ with increasing disorder strength within the explored window, indicating stronger memory effects under moderate randomness due to broadened revival windows and more irregular coherent-feedback interference.
The major peaks of $N$ are temporally aligned with revival windows in Figs.~\figpanel{F2}{c} and~\figpanel{F2}{d}. 
For $\delta=0.02$, the peak $N$ values in Fig.~\figpanel{F2}{d} exceed those in Fig.~\figpanel{F2}{c} by approximately $0.15$--$0.25$ (based on peak-value comparison in the plotted time window), showing stronger memory at larger coupling-point separation. 
Therefore, coupling-point separation acts as a direct geometrical control knob, while the disorder strength should be viewed as an effective (statistical) parameter, often set by imperfections or engineered randomness, that modulates the complexity of coherent-feedback pathways; the former mainly sets the delay timescale.

\begin{figure}[t]
\includegraphics[width=8.4 cm]{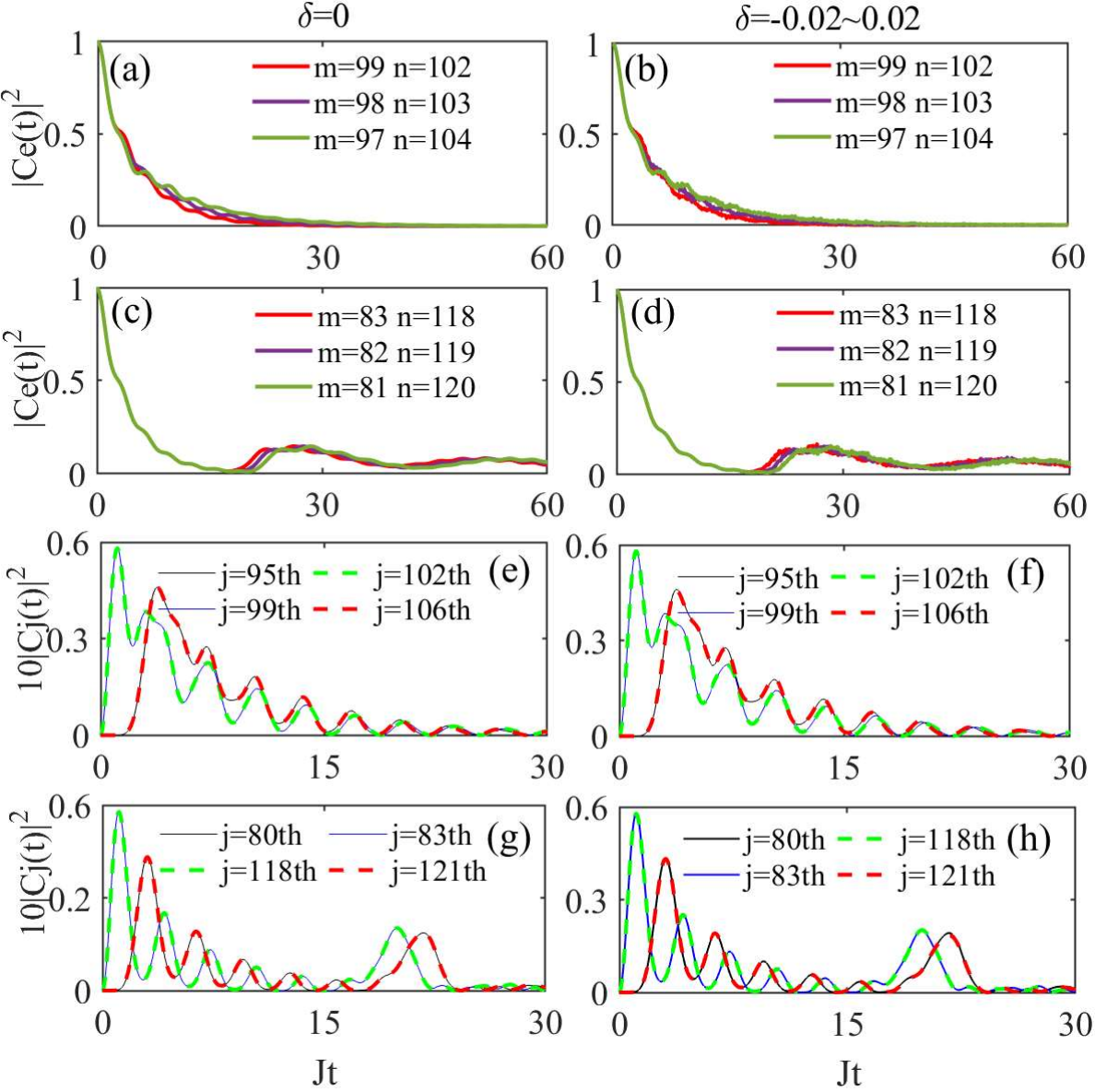}
\caption{\label{F3}
(a) and (c) Evolution of the excited-state population $|C_e(t)|^2$ in an ordered lattice ($\delta = 0$) for different coupling-point separations $|m - n|$.  
(b) and (d) Evolution of $|C_e(t)|^2$ in a disordered lattice with $\delta \in [-0.02,0.02]$ for different $|m - n|$.  
(e) and (g) Evolution of the photon population at site $j$ in an ordered lattice ($\delta = 0$).  
(f),(h) Time evolution of the photon population at site $j$ in a disordered lattice with $\delta \in [-0.02,0.02]$.  
The coupling points are $m=99$ and $n=102$ in (e) and (f), while they are $m=83$ and $n=118$ in (g) and (h).  
Other parameters are the same as in Fig.~\ref{F2}.}
\end{figure}


Figures~\figpanel{F3}{a},~\figpanel{F3}{c},~\figpanel{F3}{e}, and~\figpanel{F3}{g} illustrate the temporal evolution characteristics of the atomic population and the lattice site during the interaction between the giant atom and the ordered lattice (\(\delta = 0\)).  
Figures~\figpanel{F3}{b},~\figpanel{F3}{d},~\figpanel{F3}{f}, and~\figpanel{F3}{h} correspond to the scenario with disordered lattices (\(\delta = 0.02\)).  
For short coupling-point separation, the atomic population exhibits quasi-exponential decay behavior, and the degree of disorder has no significant modulatory effect on the evolution trend as the coupling-point separation increases.  
However, in Figs.~\figpanel{F3}{c} and~\figpanel{F3}{d}, for larger separation, the re-excitation onset shifts to later times, consistent with increased photon flight time between coupling points and delayed reabsorption.
This is attributed to the prolonged reabsorption time of delayed photons at the other coupling point due to the increased coupling-point separation.  
Comparing Figs.~\figpanel{F3}{e} with~\figpanel{F3}{f}, and Figs.~\figpanel{F3}{g} with~\figpanel{F3}{h}, it is evident that photon-population evolution across lattice sites remains globally robust under disorder, while local interference textures become rougher due to random phase accumulation.  
Outside the coupling region, the probability amplitude distribution of the lattice exhibits exponential-like decay characteristics with high symmetry. 
Between the two coupling points, disorder modifies fringe contrast and phase-sensitive interference details, evidencing quantitative sensitivity of coherent feedback pathways.

\begin{figure}[t]
\includegraphics[width=8.4 cm]{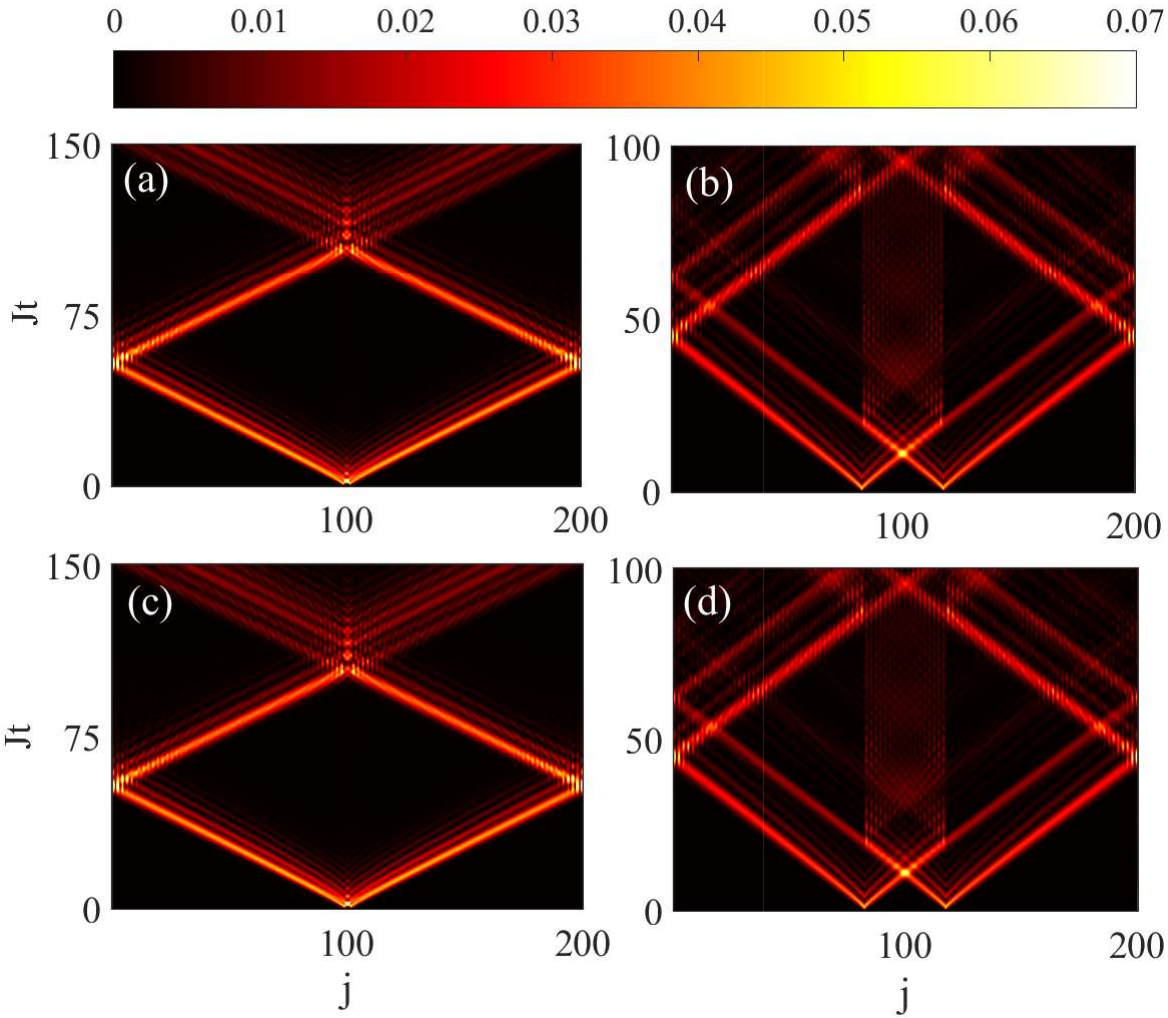}
\caption{(a) and (b) Dynamical evolution of the photon state in the ordered lattice (\(\delta = 0\)).  
(c) and (d) Dynamical evolution of the photon state in the disordered lattice ($\delta \in [-0.02,0.02]$).  
Here, \(m=99\) and \(n=102\) in (a) and (c), and \(m=83\) and \(n=118\) in (b) and (d).  
Other parameters are the same as in Fig.~\ref{F2}.}
\label{F4}
\end{figure}

Figure~\ref{F4} illustrates the dynamical evolution of photon propagation in the lattice coupled to a giant atom. 
The lattice is ordered in Figs.~\figpanel{F4}{a} and~\figpanel{F4}{b}, and disordered in Figs.~\figpanel{F4}{c} and~\figpanel{F4}{d}.  
A comparison between Figs.~\figpanel{F4}{a} (\figpanel{F4}{b}) and~\figpanel{F4}{c} (\figpanel{F4}{d}) demonstrates that photon dynamics remain robust in disordered lattices at the global level (bidirectional emission and boundary-reflection paths are preserved), while the local interference morphology is perturbed. 
As depicted in Figs.~\figpanel{F4}{a} and~\figpanel{F4}{c}, excited photons originating from the giant atom are injected into the discrete lattice system from the coupling points \(m= 99 \) and \( n = 102 \), respectively. 
Photons emitted from these two sites \( m = 99 \) and \( n = 102 \) propagate bidirectionally and undergo boundary reflections with symmetric trajectories, confirming stable transport channels.
Figures~\figpanel{F4}{b} and~\figpanel{F4}{d} follow the same propagation principles, but notable interference fringes are observable between the coupling points, with two distinct reflection sites appearing at the system boundaries.

\subsection{The influence of disorder on energy spectrum}
\label{SecIV}

\begin{figure}[t]
\includegraphics[width=8.4cm]{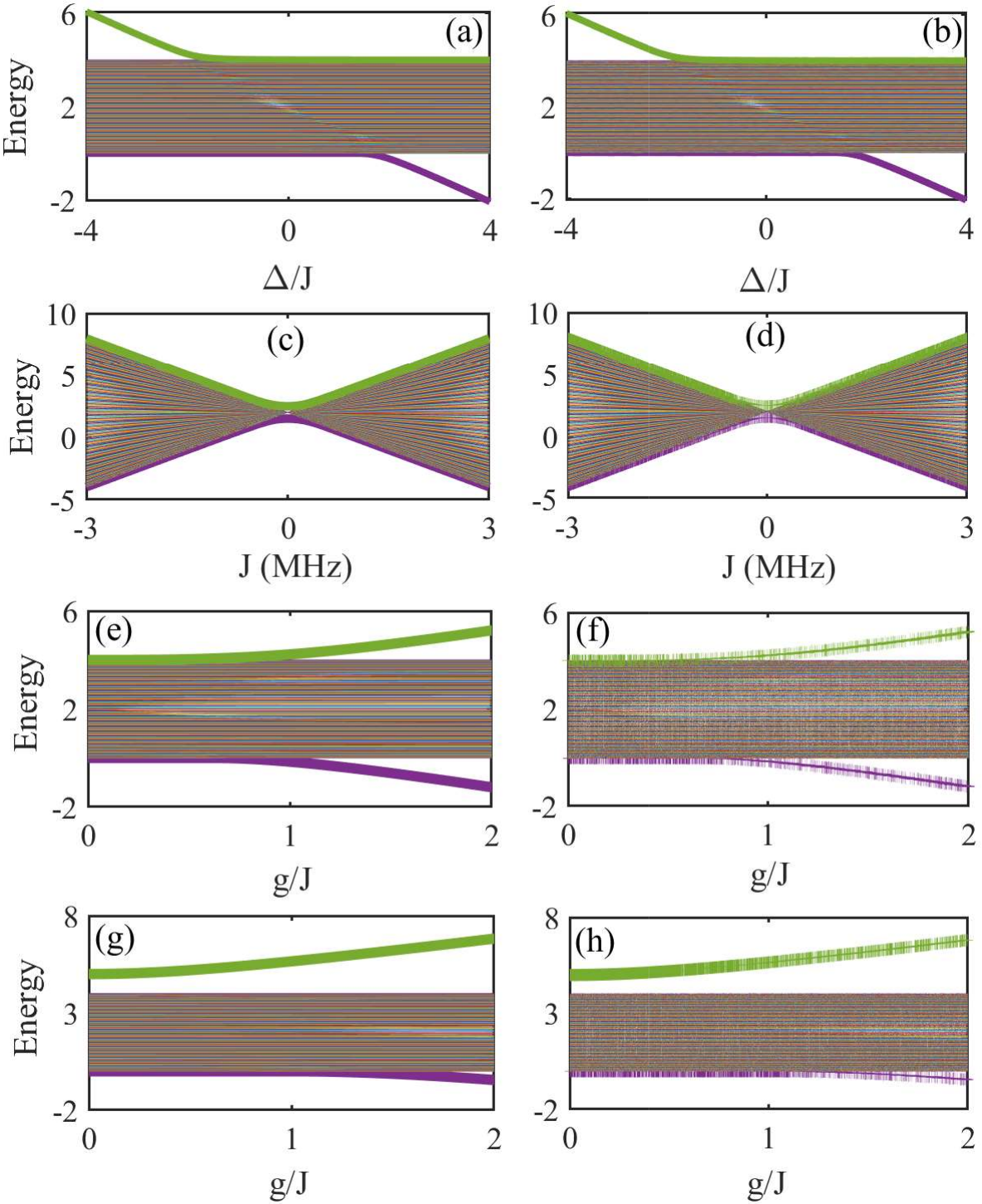}
\caption{(a) and (b) Energy spectrum as a function of detuning $\Delta = \omega_{0} - \omega_{e}$.  
(c) and (d) Energy spectrum as a function of hopping strength $J$.  
(e) and (f) Energy spectrum as a function of coupling strength $g$.  
(g) and (h) Energy spectrum as a function of coupling strength $g$ with $\omega_{0} = 2J$ and $\omega_{e} = 5J$.  
Here, $\delta = 0$ for (a), (c), (e), and (g), and $\delta \in [-0.02,0.02]$ for (b), (d), (f), and (h).  
Here, $m=99$ and $n=102$, and other parameters are the same as in Fig.~\ref{F2}.
\label{F5}}
 \end{figure}

To reveal how lattice disorder reshapes the energy spectrum, we compare the spectra of the ordered and disordered lattices under identical parameters. We choose parameters that place excitations in the band-gap region so that localized features and bound-state branches are visible, which allows us to directly identify the disorder sensitivity of these spectral components.
Figure~\ref{F5} compares the energy spectra of a lattice coupled to a giant atom as functions of detuning, hopping strength, and coupling strength for the ordered case and the disordered case. The overall scattering-band structure is largely preserved under disorder, whereas the out-of-band branches associated with bound states exhibit visible fluctuations and broadening, indicating stronger disorder sensitivity.

As shown in Figs.~\figpanel{F5}{a} and~\figpanel{F5}{b}, the $4J$-wide central band corresponds to propagating scattering states, while the upper/lower out-of-band branches indicate localized bound states around coupling points.
In the disordered spectrum, the central scattering band remains nearly unchanged in its entirety, while the bound-state branches become jagged and slightly shifted, reflecting disorder-induced spectral fluctuations.
In Figs.~\figpanel{F5}{c} and~\figpanel{F5}{d}, localized states emerge when the hopping strength \(J\) falls within the interval \([-0.1, 0.1]\). 
Disorder does not qualitatively alter the condition for the emergence of localized features, but it introduces noticeable irregularities and parameter-dependent fluctuations in the bound-state-related branches.
For \(\omega_{0} = 2J\) and \(\omega_{e} = 5J\) in Figs.~\figpanel{F5}{e} and~\figpanel{F5}{f}, bound-state characteristics become evident for sufficiently strong coupling (e.g., \(g > J\) in the displayed examples).  
We find that the impact of disorder is manifested primarily as fluctuations and broadening of the bound-state branches, whereas the scattering-band background remains comparatively stable.
Combining the results from Figs.~\figpanel{F5}{a} and~\figpanel{F5}{b}, photon localization occurs when \(|\Delta| > 2J\), and bound states persist across the entire parameter range of coupling strength \(g\).
This spectral dichotomy provides a consistent interpretation of the time-domain results.
The robustness of the scattering-band framework under disorder supports the observed stability of the overall decay envelope and global transport patterns, while disorder-sensitive bound-state branches and localization features reshape revival windows and facilitate enhanced information backflow.

\section{Conclusion}
\label{SecV}

In this work, we have developed a disorder-aware giant-atom lattice model with random on-site frequency fluctuations and performed a unified analysis in the time domain and in the energy spectrum. 
We found that, across the explored moderate disorder range, the overall relaxation envelope and global photon-transport patterns remain robust against disorder, indicating stable dynamics in the presence of lattice imperfections.
Meanwhile, the normalized non-Markovianity can increase markedly with disorder strength, showing that moderate lattice disorder can enhance delayed information backflow rather than merely suppress coherent effects. 
This trend is consistent with the disorder-induced broadening and irregularization of revival windows in the excited-state amplitude.
The underlying mechanism can be understood from the spectral structure.
Scattering-band transport remains comparatively robust, while disorder strongly perturbs the bound-state branches and enhances localization features that enrich coherent-feedback pathways.
The coupling-point separation primarily sets the feedback delay timescale, whereas disorder modulates the interference complexity of the returning field. 
These results clarify how the robustness of the overall decay envelope can coexist with enhanced non-Markovian memory in a disordered lattice reservoir.
They also provide theoretical guidance for designing defect-tolerant giant-atom quantum devices whose non-Markovian memory is tunable through geometrical control of the coupling-point separation and, in platforms where it is feasible, through statistically engineered on-site detunings.

\section*{Acknowledgments}
This work is supported by the Science Foundation of the Education Department of Jilin Province (No. JJKH20250301KJ) and Jilin Scientific and Technological Development Program (No. 20240101328JC).



\vspace{8pt}
\end{document}